\documentstyle[multicol,aps,prl,psfig,epsf]{revtex}

\newcommand{\beq}{\begin{equation}}
\newcommand{\eeq}{\end{equation}}
\newcommand{\bdis}{\begin{displaymath}}
\newcommand{\edis}{\end{displaymath}}
\newcommand{\bea}{\begin{eqnarray}}
\newcommand{\eea}{\end{eqnarray}}
\newcommand{\barr}{\begin{array}}
\newcommand{\earr}{\end{array}}

{}
\begin{document}
\twocolumn[\hsize\textwidth\columnwidth\hsize\csname 
 @twocolumnfalse\endcsname

\title{Coarsening and Slow-Dynamics in Granular Compaction}
\author{
A. Baldassarri$^{1}$, S. Krishnamurthy$^{2}$, V. Loreto$^{3}$ and S. Roux$^{4}$}

\address{
$^1$ INFM  UdR Camerino, Universit\`a di Camerino, Via Madonna delle
Carceri I-62032 Camerino, Italy \\
$^2$ Department of Theoretical Physics, University of Oxford, 1 Keble Road,
Oxford OX1 3NP, UK\\
$^3$ Universit\`a degli Studi di Roma ``La Sapienza'', 
P.le A. Moro 5, 00185 Rome, Italy and INFM, Unit\`a di Roma 1\\
$^4$ Laboratoire Surface du Verre et Interfaces,
Unit\'e Mixte de Recherche CNRS/Saint-Gobain,\\
 39, Quai Lucien Lefranc, F-93303 Aubervilliers Cedex, France.
}
 
\maketitle

\date{\today}
\begin{abstract}
We address the problem of the microscopic reorganization 
of a granular medium under  a compaction process in the framework of
{\em Tetris}-like models. We point out the existence of regions of spatial
organization which we call {\it domains}, and study their time
evolution. It turns out that after an initial transient,
most of the activity of the system is concentrated on the boundaries between
domains. One can then describe the compaction phenomenon as a coarsening 
process for the domains, and a progressive reduction of domain
boundaries. We discuss the link between the coarsening process and the slow
dynamics in the framework of a model of active walkers on 
active substrates.
\end{abstract}
\smallskip
\vskip2pc]

The phenomenon of granular compaction involves the increase of the
density of a granular medium\cite{grain} subject to shaking or
tapping. Triggered by experimental results of the Chicago
group\cite{Knight}, that suggested that compaction follows an inverse
logarithmic law with the tapping number, several models have been
proposed to explain the slow relaxation features of granular media
\cite{BarkerMehta,Ben-Naim,degennes,prltetris,NCH,brey,head,satya}.
Though in all these different cases a very slow relaxation
(eventually logarithmic) is reproduced, an explicit connection between
the above models and a {\it real} granular medium is however still
rather tenuous.

The aim of this paper is to elucidate the origin of the 
very slow relaxation studying explicitly the microscopic response 
of a granular medium subject to shaking.
We address this problem within the framework of the recently
introduced class of {Tetris}-like models \cite{prltetris}
which are known to reproduce several features observed experimentally
in granular  materials such as slow dynamics, segregation, aging and
hysteresis.  

We find, quite surprisingly, that the system reorganizes under the
shaking dynamics into several ordered regions (see
\cite{segtet,struct} for other examples).
We call these {\it domains}, and study their
time evolution.  After a short transient, most of the activity
of the system is concentrated along the boundaries between domains
(we note that this concerns only a small part of the entire system).
Under shaking, the domain boundaries move throughout the system and 
free the vacancies they encounter leading to a progressive
densification.  Moreover, when two domain boundaries meet, they annihilate.
One can thus describe the compaction phenomenon in this system, 
as a coarsening \cite{coarsen} (i.e. a domain growth) process.
As the system compactifies, the domains coarsen and the boundary regions
are reduced and thus the process becomes slower.
We give a quantitative description of this phenomenon 
studying the behavior of the space-time correlation function that 
is expressed by $C(r,t) \simeq f\left(r/\xi(t)\right)$
where $\xi(t)$ is the correlation length, say the typical size
of a domain.
This coarsening of domains is related to the slow compaction process
by measuring 
the {\em persistence} \cite{persistence} exponent of the 
phenomenon, as well as by measuring the activity and the 
motion of the domain boundaries.

Let us briefly recall the definition of the {\em Tetris}
model. Although this class of models allows for an infinity of
particles types, shapes and sizes, here we use,
without loss of generality for the main features, a system of
elongated particles. These occupy the sites of a square lattice tilted
by $45^{\circ}$, with periodic boundary conditions in the horizontal
direction (cylindrical geometry) and a rigid plane at its bottom. 
Particles cannot overlap and this
condition produces strong constraints (frustration) on their
relative positions.  The system is initialized by inserting grains at
the top of the system, one at the time, and letting them fall down,
performing under the effect of gravity, an oriented random walk on the
lattice, until they reach a stable position, i.e. a position from
which they cannot fall further.  The effect of vibrations is
implemented by means of a two-steps Monte-Carlo algorithm mimicking a
tapping procedure. The role of the tapping amplitude is played by a
parameter $0 < x < 1$ that describes the strength of the bias in the
particle movement, induced by the gravity (we refer to
\cite{prltetris} for the details).

\begin{figure}[tbp]
\centerline{
        \psfig{figure=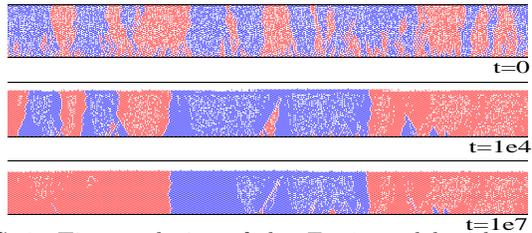,width=7cm,height=3cm,angle=0}}
\caption{Time-evolution of the {\em Tetris} model with $x=0.5$
resolved in antiferromagnetic domains at times (number of taps) $t=0$,
$10^4$ and $10^7$ from top to bottom.  White squares are voids while
red and blue squares represent particles belonging to the two possible
domain types.}
\label{cluster1}
\end{figure}


In the simplest version, the {\em Tetris} model consists of a single
rod-like type of particle (rectangles of uniform size $a \times b$
with $a=0.75$ mesh units and $b=0$) with two possible orientations
(along the principal axis of the lattice) chosen to be equally
probable. A generic configuration can be described by assigning to
each site of the lattice $(x,y)$ ($0<x<L_x$ is the horizontal
coordinate and $0<y<L_y$ is the vertical one), a variable
$\sigma(x,y,t)$ whose value is $0$ if the site is empty and $\pm 1$ if
the site is occupied by a particle with one of the two possible
orientations. At every density the system can be resolved into
domains, i.e. regions in which the staggered magnetization keeps a
definite sign and each domain presents an antiferromagnetic order with
vacancies ($+1$ particles on odd (even) rows and $-1$ particles on
even (odd) rows).  One can then observe the evolution of the
compaction dynamics in terms of the evolution of the domains.

At the beginning, after pouring the grains into the container, 
i.e. in the so-called loose density state, the system presents 
a disordered structure with an alternation of the two types of domains,
even though the domain boundaries do not as yet span the system
from top to bottom.
At this stage the number of domain boundaries depends on the aspect ratio
(height by width) of the container: the smaller  the aspect ratio
(wider is the system) the larger  the number of domains. 
The domain size is of the order of the height of the system, and is 
almost independent of the horizontal size of the system $L_x$, 
as long $L_x>L_y$. 

The compaction can now be seen as a slow elimination of the voids
frozen in the different domains. Since the system changes only at the
domain boundaries, a void in the bulk of the domain can be freed only
when it comes in contact with a domain boundary. The domain boundaries
are then the only regions where the activity of the system is
concentrated.  Fig.\ref{cluster1} shows an example of the
time-evolution of the {\em Tetris} model resolved in antiferromagnetic
domains~\cite{animation}.  It is important
to note how narrow systems ($L_x\ll L_y$) may display a pathological
behavior (blocking) if the system has an almost single-domain like packing.

Let us now describe the coarsening dynamics in a 
more quantitative way.
We have monitored the evolution of the (longitudinal) correlation function
defined as:
\beq 
C(r,t)=\frac 1{N_{p}} \sum_{y=0}^{L_y/2} 
\sum_{x=0}^{L_x} \sigma(x,y,t)\sigma(x+r,y,t)
\label{corrdef}
\eeq 
where $N_p$ is the number of particles in the bottom half of the 
system. 
A pair of particles inside the same domain gives a positive contribution
to $C(r,t)$ while a pair in different domains gives a 
negative contribution. With this definition, the correlation function
is not sensitive to density changes and  reflects only 
the evolution of the domain sizes.

We perform extensive simulations of wide systems (to avoid blocking)
and we try a scaling collapse of several $C(r,t)$ curves at different
times: i.e. we look for a characteristic length $\xi(t)$ such that
$C(r,t)=f(r/\xi(t))$, the length $\xi(t)$ representing the average
(horizontal) domain size. As in standard coarsening dynamics, 
$\xi(t)$ grows in time and when $\xi(t) \approx
L_x$ the growth stops (blocking for single domain systems). 

\begin{figure}[t]
\centerline{
        \psfig{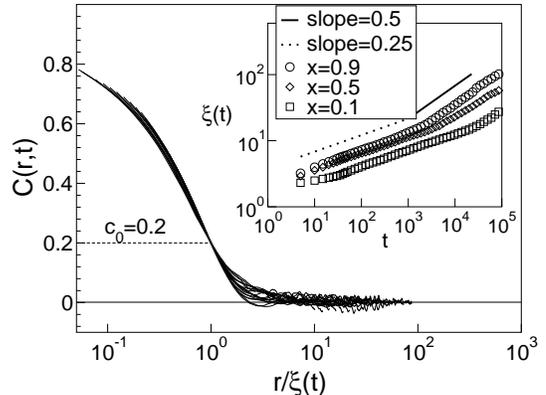}}
\caption{Scaling collapse of (1) for a system with $L_x=800$ and
$L_y=50$, simulated up to a time $t=10^5$ with $x=0.1$, and averaged
over $200$ different dynamics.  $\xi(t)$ is chosen such that
$C(\xi(t),t) = C_0$ for all the curves obtained at different times for
$C_0 =0.2$ (One gets the same result for a large range of values of
$C_0$). In the inset is shown $\xi(t)$ vs.  $t$ for different values
of $x=0.1,0.5,0.9$. The crossover occurs at an $x$-dependent $t^*$
such that $\xi(t^*) \simeq L_y$.  In the second regime one observes an
exponent equal to $0.5$ only for $x=0.9$ while slightly smaller
exponents are observed for smaller $x$. We believe these deviations
from an exponent $0.5$ are transients evolving towards the asymptotic
diffusive value. }
\label{cross}
\end{figure}

However, the height, $L_y$, of the system is another characteristic
scale, and depending on whether $\xi$ is smaller or larger than $L_y$,
two different  regimes can be observed.
A quantitative analysis of $\xi$ versus time reveals that 
for $\xi(t)< L_y$, $\xi(t)\propto t^{0.25}$, whereas for $\xi(t)>L_y$, 
we observe a faster growth : $\xi(t)\propto t^{0.5}$ (See Fig.\ref{cross}).

These results can be interpreted according to the following scenario.
One can imagine in general that the size $S(t)$ (area) of the domains
grows as  
a power of time. The correlation length $\xi(t)$ is a 
measure of the lateral size of the domains. Now, as long as the domains
grow in an isotropic way we can expect the area to scale as $S(t)
\sim \xi(t)^2$. 
In this regime the domain walls, though biased by gravity, do not
yet span the system in the $y$ direction. 
Later on, when the domain walls span the system from top to bottom,
the coarsening dynamics is dominated by the diffusion of these almost
vertical walls, which eventually collide and annihilate each other.
At this stage we expect the area scales as $S(t) \sim L_y \xi(t)$. 
The crossover we observe is then compatible with a growth $S(t) \sim
t^{\frac{1}{2}}$ 
which gives $\xi(t) \sim t^{1/4}$ in the early regime followed by the
asymptotic  
behaviour $\xi(t) \sim t^{1/2}$. The crossover between the first and the second regime
is evident in the first two pictures of Fig.\ref{cluster1}.

\begin{figure}[tp]
\centerline{
        \psfig{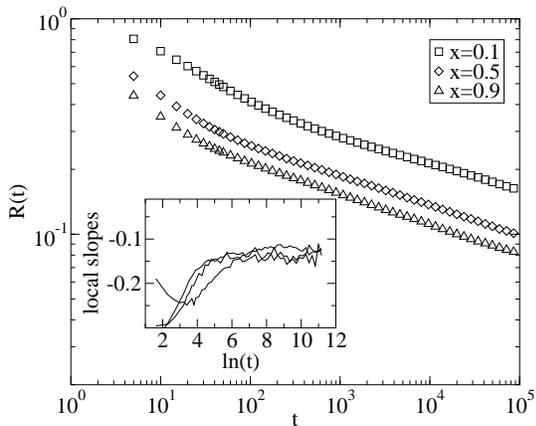}}
\caption{$R(t)$ (see text for the definition) vs. time for different values 
of $x$. The inset shows the local logarithmic slopes.}
\label{persistence}
\end{figure}

How now do we relate this coarsening behaviour to the density relaxation?
The motion of the domain walls occurs in a background of vacancies. 
We observe that regions swept through by the domain walls compactify
by triggering  particle rearrangements, while regions not yet swept through
remain disordered. Thus within each domain there is a compactified
region through which the domain wall has made several forays, and a
disordered loose region  through which the domain wall has not yet
swept through.
The compaction process is related to the
growth of the dense  ordered regions rather than to the characteristic
length of the domains. 
The fast dynamics of the domain walls is thus not in contrast 
with the slow dynamics of the bulk density.

\begin{figure}[tp]
\centerline{
        \psfig{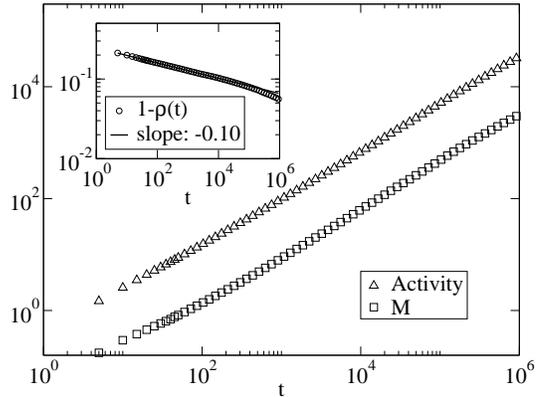}}
\caption{
$A(t)$ and $M(t)$ (see text for the definition) 
vs. $t$ for $x=0.5$, $L_x=800$, $L_y=50$. Similar results have been obtained
with $x=0.1$ and $x=0.9$. The inset reports $(1-\rho(t))$ vs. $t$~[15].}
\label{activity}
\end{figure}

In order to support this picture we have measured the fraction of 
persistent sites (or persistence probability) $R(t)$ \cite{persistence}, 
i.e. the fraction of sites that never changed their status up to time $t$.
If the triggering process mentioned above was perfectly efficient,
i.e. every time a domain wall passes through a vacancy one triggers 
a process increasing the density locally, one would expect that
$1 -\rho(t)$ is described by $R(t)$. Otherwise the behaviour of 
$1 -\rho(t)$ is slower than $R(t)$.
In analogy with recent studies \cite{persistence} in standard coarsening models, 
a very slow algebraic decrease of such quantities is observed as a function of time;
we obtain in particular  that $R(t)$ (see Fig.\ref{persistence})
scales as $t^{-\theta}$ with $\theta \simeq 0.15$.
The behaviour of $(1-\rho(t))$ vs. $t$ is shown in the inset of Fig.\ref{activity}.
We observe that it is consistent with an algebraic decay
$1-\rho(t)\simeq t^{-0.10}$ with an exponent 
smaller than the persistence exponent $\theta$ which is compatible
with our discussion above.

Additional information on the system can be obtained by monitoring the
following quantities. (1) $A(t)$: activity in the system measured as
the cumulative number of successful moves; (2) $M(t)$ is intended to
measure the mobility of the domain walls. It is obtained by the
cumulative sum of the absolute value of the derivative of the
staggered magnetization in thin vertical stripes, averaged over all
the stripes. From Fig.\ref{activity}, which reports the results for
both quantities, one deduces two main things: the activity of the
system is concentrated on the domain walls (since $A(t)$ and $M(t)$
have the same functional form up to a constant) and both $A(t)$ and
$M(t)$ scale as $(1-\rho(t))^{-\beta}$.

In order to get a better insight into the above  mentioned phenomenology
it would be  very interesting to have a quantitative understanding  of
the link between the coarsening process and the very slow global density
relaxation. Under the hypothesis that the system is translationally
invariant  in the direction of gravity, the medium is described as a one
dimensional ($y$-averaged) density profile, $\rho(x,t)$, in which 
particles (the domain wall) move.  To comply with
our previous description, we assume that the density $\rho$ is only
susceptible to increase at the positions occupied by the walkers, and
remains quenched elsewhere. Moreover, in order to follow more closely 
what happens in the {\em Tetris} model we consider a situation where 
the motion of the walkers is coupled to the environment, i.e. 
the local density via  a potential field depending only on the density.
The problem can be cast in terms of two coupled 
equations: one describing the over-damped motion of the walkers and
another describing the evolution of the local density as:
\begin{eqnarray}
\frac{dX}{dt} & = &  
- \frac {\partial V[\rho(x,t)]}{\partial \rho(x,t)} \bigg{\vert}_{x=X(t)} 
 \frac {\partial \rho(x,t)}{\partial x} \bigg{\vert}_{x=X(t)}
+  \Gamma (t) \nonumber \\
\frac{\partial \rho(x,t)}{\partial t} & = & \,\, f(\rho(x,t)) \,\, \delta(x-X(t))
\label{model-1d}
\end{eqnarray}
where $\Gamma(t)$ is an uncorrelated Langevin force with
$\langle\Gamma (t)\rangle = 0$ and $\langle\Gamma(t)
\Gamma(t^{\prime})\rangle = q \delta(t-t^{\prime})$.  The potential
$V$ attracts the walker to regions where activity has been intense,
and repels it from unvisited regions.

A detailed treatment of the above defined equations 
is presented elsewhere \cite{langevin}. 
Here we summarize the main features. 
The function $f$ should be such that $\rho$ approaches unity for long
times, and hence $f(\rho)\to 0$ as $\rho\to 1$.  An example of such an
$f$ is $f(\rho)=(1-\rho)^a$, where $a$ is a positive exponent.
The potential $V$ should provide a drift toward high density regions.  A
simple suitable form is $V(\rho)=-\rho^{\gamma}$.  In fact it is possible to
show that the latter functional form is inessential, provided $V(\rho)$ 
behaves linearly in $\rho$ as $\rho \rightarrow 1$. In this way 
$\frac{\partial V[\rho(x,t)]}{\partial \rho(x,t)}$ becomes a constant,
i.e. unimportant, as $\rho$ tends to $1$.
With these definitions eqs.(\ref{model-1d}) can be recast in the form
\begin{equation}
\frac{dX}{dt}= F^{-1-b} \nabla F +\Gamma(t), \;\;\;
\frac{dF}{dt}=\delta(x-X(t))
\label{recast}
\end{equation}
where $b = 1/(a-1)$ and we have introduced the function 
$F=[1/(a-1)](1-\rho)^{1-a}$ which, according to Eq.~\ref{recast}, 
represents the cumulative activity on the site $x$ (in general $dF/d\rho=\/f(\rho)$).
In this way the choice of the function $f$ is consistent
with the results obtained for $A(t)$ and $M(t)$ in the Tetris model.
The main idea behind this kind of modeling is that
the high density regions (i.e. the potential wells) 
tend to trap the walkers, that, in their turn 
are able to change the environment, 
i.e. the local density, though their efficiency decreases with the 
increase of the density.
From the combination of these two effects a drastic slowing down is expected.
The way the walkers escape from the potential wells is to progressively
carve their way out by pushing the potential barrier and so enlarging the
compactified region.

Different aspects come into play in the compaction process.  One of
them is related to the fact that initially a large number of walkers
is to be introduced.  Since our modeling only concerns the regime
where a one dimensional description is adapted, the typical distance
between walkers is proportional to $L_y$.  However, when two walkers
meet, they annihilate, and thus the subsequent increase the density
becomes less and less effective.  The quantification of this effect
has been done in a previous section, through the pair correlation
function of the domains.  By itself, this single aspect is not
sufficient to account for the slow densification observed numerically.
The second aspect concerns the densification due to a single walker.
Starting from a low density for the medium, we observe that the
density does not remains uniform.  Starting from any site, at low
temperature, the domain walls first drill a potential well where they
lie.  However, as the density approaches unity, the densification
becomes less and less efficient.  The only option for the walkers is
to expand in lateral size through a progressive translation of the
well boundaries.  Specific solutions of this regime can be obtained as
solitary waves~\cite{langevin}.  In this second regime, the
densification rate is controlled by the velocity of the
latters. Finally, the wells tend to coalesce and the mean density
decays as $1-\rho\sim t^{-1/(a-1)}$.  Though the variety of the
different phenomena involved in the density evolution, (number of
walkers, width and depth of potential wells, late stages crossover
phenomena) renders difficult the identification of the $F$ function in
Eq.3, we retain that the main features observed in the Tetris model
are captured by this walkers modeling.  

It is also important to stress that the equivalence observed between
the activity and the motion of the domain boundaries, implies that the
same treatment could also be carried out for particles with random
shapes\cite{prltetris}. Here the existence of domains is no longer
evident but the activity still remains confined and it is not spread
out uniformly over the system.  Finally it is worth to remark as our
analysis could be easily exported in experimental setup like the one
proposed in \cite{olson}.  {\large Acknowledgments} We thank
H.J.Herrmann for useful discussions. This work has been partially
supported from the European Network-Fractals under contract
No. FMRXCT980183 and by INFM {\em Center for Statistical Mechanics and
Complexity} (SMC) and PAIS Project (AB).  SK would like to thank
EPSRC, UK for a research fellowship.


\begin{thebibliography}{99}
\bibitem{however} However, we have indication of a
second crossover in the temporal behaviour of $R(t)$ and $\rho(t)$, at
very large times.  In particular $(1-\rho(t))$ seems to undergo an
acceleration.
\end{thebibliography}
\end{document}